\documentclass{amsart}
\usepackage[cp1251]{inputenc}
\usepackage[english]{babel}
\newtheorem{theorem}{Theorem}
\newtheorem{definition}{Definition}
\newtheorem{lemma}{Lemma}
\usepackage{amsmath}
\usepackage{amssymb}
\usepackage{amsfonts}

\begin{document}

\title[EQUIVALENCE CLASSES OF THE SECOND ORDER ODEs \dots]{EQUIVALENCE CLASSES OF THE SECOND ORDER ODEs WITH THE CONSTANT CARTAN INVARIANT}

\author{VERA V. KARTAK}

\maketitle

{
\small
\begin{quote}
{{\bf Address.} Chair of Higher Algebra and Geometry, Bashkir State University, ul. Z.Validi, 32, Ufa, 450074, Russia.\\ {\bf E-mail:} kvera@mail.ru}
\medskip

\noindent {\bf Abstact.} Second order ordinary differential equations that possesses the constant invariant are investigated.
Four basic types of these equations were found. For every type the complete list of nonequivalent equations is issued. As the exampes the equivalence problem for the Painleve II equation, Painleve III equation with three zero parameters, Emden equations and for some other equations is solved.

\medskip
 

\medskip

\end{quote}
}

\section{Introduction}

Let us consider the following second order ODE:
\begin{equation}\label{eq}
y''=P(x,y)+3\,Q(x,y)y'+3\,R(x,y)y^{\prime 2}+S(x,y)y^{\prime 3}.
\end{equation}
General point transformations
\begin{equation}\label{zam}
\tilde x=\tilde x(x,y),\quad \tilde y=\tilde y(x,y)
\end{equation}
preserve  the form of equation (\ref{eq}). 

Let us consider two arbitrary equations (\ref{eq}). The problem of
existence of the point transformation (\ref{zam}) that connects these equations is called {\it the Equivalence Problem}.
For the arbitrary equations (\ref{eq}) the explicit solution of the equivalence problem is rather complicated,  see \cite{Tresse1}, \cite{Tresse2}.
It was effectively solved for the linear equations, see  \cite {Liouville}, \cite{Grissom}, \cite{Ibragimov}, \cite{Yumaguzhin}, \cite{Sharipov2}, \cite{Babich};
for some Painleve equations, see \cite{Lamb},  \cite{Sokolov}, \cite{Babich},\cite{Hietarinta}, \cite{Dridi}, \cite{Kartak}; for the Emden equation, see \cite{BordagBandle} and for other equations, for example, see \cite{Sharipov2}, \cite{Meleshko}.

The main approach that allows to solve the equivalence problem is based on the Invariant Theory.
{\it Invariant}  is a certain function depending on $(x,\,y)$ that is unchanged under the transformation (\ref{zam}):  $I(x,y)= I(\tilde x (x,y),\tilde y(x,y)).$

{\it Pseudoinvariant  of weight $ m $}  is a certain function depending on $(x,\,y)$ that is transformed under (\ref{zam}) with  factor $\det T$ (the Jacobi determinant) in the degree $m$: 
$$J(x,y)=(\det T)^m \cdot  J(\tilde x (x,y),\tilde y(x,y)),\quad
T=\left(\begin{array}{cc}
	 \partial \tilde x /\partial  x& \partial \tilde x /\partial  y\\ \partial \tilde y /\partial  x & \partial \tilde y /\partial  y
	 \end{array}\right).
$$

{ \it Pseudotensorial field of weight $ m $ and  valence $ (r, s) $} is an indexed set that  transforms  under change of variables (\ref{zam}) by the rule
$$
F^{i_1\dots i_r}_{j_1\dots j_s}=(\det T)^m{\sum_{p_1\dots p_r}}
{\sum_{q_1\dots q_s}} S^{i_1}_{p_1}\dots S^{i_r}_{p_r}T^{q_1}_{j_1}
\dots T^{q_s}_{j_s} \tilde F^{p_1\dots p_r}_{q_1\dots q_s},
$$
here $S=T^{-1}$. It is easy to check that only factor $ (\det T)^ m $  distinguishes 
the pseudotensorial field from the classical  tensorial field.

Invariant Theory of equation (\ref{eq}) goes back to the classical works of R.Liouville \cite {Liouville}, S.Lie \cite{Lie}, A.Tresse \cite{Tresse1}, \cite{Tresse2}, E.Cartan \cite{Cartan}, \cite{Thomsen} (Late 19th- and Early 20th-Century) and 
continues in the works of \cite{Grissom}, \cite{Lamb}, 
\cite{Hietarinta}, \cite{BordagDruima}, \cite{Babich}, \cite{Sharipov1}, \cite{Sharipov2}, \cite{Sharipov3}, \cite{Ibragimov} (Late 20th-Century).  
It remains an active research topic in the  21th-Century, see  \cite{BordagBandle}, \cite{Kartak}, \cite{Meleshko}. 
Background is adequately described  in papers \cite{Babich}, \cite{BordagBandle}.

In the present paper we use notations from works \cite{Sharipov1}, \cite{Sharipov2}, \cite{Sharipov3}, \cite{Kartak}  to calculate the invariants and pseudoinvariants 
of equations (\ref {eq}). The correlation between these (pseudo)invariants and semiinvariants from works \cite{Cartan}, \cite{Liouville} (as they were presented in  \cite{BordagBandle}) shows in the next chapter. 
The explicit formulas for their computation via known functions
$P(x,y)$, $Q(x,y)$, $R(x,y),$ $S(x,y)$ 
contained in the Appendix A. Here and everywhere below  notation $K_{i.j}$ denotes the partial differentiation:   $K_{i.j}={\partial ^{i+j}K}/{\partial x^i\partial y^j}.$ 
  
  \section{Computation of invariants}
  \subsection{Some geometric objects}
  
 {\bf Step 1.} From the functions $P$, $Q$, $R$ and $S$, that are the coefficients of equation (\ref{eq}),  let us organize the 3-indexes massive by the following rule:
$$
\aligned
\Theta_{111}&=P, \qquad \Theta_{121}=\Theta_{211}=\Theta_{112}=Q,\\
\Theta_{222}&=S,\qquad \Theta_{122}=\Theta_{212}=\Theta_{221}=R.
\endaligned
$$
As the 'Gramian matrixes' let us take the following couple:
$$
\aligned
d^{ij}=&\left\|\begin{array}{rr}  0 & 1\\ -1 & 0
\end{array}\right\|,\quad\text{pseudotensorial field of the weight 1},\\
d_{ij}=&\left\|\begin{array}{rr}  0 & 1\\ -1 & 0
\end{array}\right\|,\quad\text{pseudotensorial field of the weight -1}.
\endaligned
$$
 {\bf Step 2.} Let us reise the first index  
\begin{equation}\label{Theta1}{\Theta_{ij}^k}=\sum_{r=1}^2d^{kr}\Theta_{rij}.
\end{equation} 
Under the change of variables  (\ref{zam})   $\Theta_{ij}^k$ transforms ``almost'' as a affine connection. (The transformation rule is into the paper \cite{Sharipov1}).

{\bf Step 3.} Using ${\Theta_{ij}^k}$ as the affine connection let us construct the ``curvature tensor'':
$$
{\Omega_{rij}^k}=\frac{\partial\Theta^k_{jr}}{\partial u^i}-\frac{\partial \Theta^k_{ir}}{\partial u^j}+
\sum_{q=1}^2\Theta_{iq}^k\Theta_{jr}^q-\sum_{q=1}^2\Theta^k_{jq}\Theta^q_{ir},\quad\text{here}\; u^1=x,\, u^2=y,
$$
and the ``Ricci tensor'' $\Omega_{rj}=\sum_{k=1}^2\Omega^k_{rkj}.$
The both objcts are not the tensors. (See \cite{Sharipov1}).

{\bf Step 4.} The following 3 indexes massive is the tensor:
$$
 {W_{ijk}}=\nabla_i\Omega_{jk}-
\nabla_j\Omega_{ik}.$$
Here we use
$\Theta_{ij}^k$ instead of  the affine connection when made the covariant differentiation.

{\bf Step 5.} Using the tensor $W_{ijk}$ let us construct the new pseudovectorial fields:
$$
\aligned
&\alpha_k=\frac 12\sum_{i=1}^2\sum_{j=1}^2W_{ijk}d^{ij}\quad \phantom{.............}\text{pseudocovectorial field of weight 1,}\\
&\beta_i=3\nabla_i\alpha_kd^{kr}\alpha_r+\nabla_r\alpha_kd^{kr}\alpha_i\quad \text{pseudocovectorial field of weight 3.}
\endaligned $$
The coincident pseudovectorial fields are:  $\alpha^j=d^{jk}\alpha_k$ of weight $2$, 
$\beta^j=d^{ji}\beta_i$ of weight $4$.

There are only 3 situations:
\begin{enumerate}
 \item Pseudovectorial field {\boldmath $\alpha$}=0, { \it maximal degeneration case}, equation  is equivalent to $ y''=0;$
  \item Fields {\boldmath $\alpha$} and {\boldmath $\beta$} are collinear: $3F^5=\alpha^i\beta_i=0$, {\it intermediate  degeneration case};
\item Fields {\boldmath $\alpha$} and {\boldmath $\beta$} are non-collinear: $3F^5=\alpha^i\beta_i\ne 0$, {\it general case}.
\end{enumerate}

At the present paper we consider the {intermediate degeneration case:} $F=0$ but  {\boldmath $\alpha$}$\ne 0.$ 

{\bf Step 6.} Let us denote the quantities $\varphi_1$ and $\varphi_2$ (their explicit formulas the are into Appendix A) and organize the affine connection $\Gamma^k_{ij}$ and the pseudoinvariant $\Omega$ of weight 1
$$
{\Gamma^k_{ij}}=\Theta^k_{ij}-\frac{\varphi_k\delta^k_j+\varphi_k\delta^k_i}{3},\qquad
{\Omega}=\frac 53\left( \frac{\partial\varphi_1}{\partial y}-\frac{\partial\varphi_2}{\partial x}
\right).
$$
The rule of covariant differentiation of the pseudotensorial field was presented in \cite{Sharipov1}: 
$$
\nabla_kF^{i_1\dots i_r}_{j_1\dots j_s}= \frac{\partial F^{i_1\dots i_r}_{j_1\dots j_s}}{\partial u^k}+
\sum_{n=1}^r\sum_{v_n=1}^2\Gamma_{k v_n}^{i_n}F^{i_1\dots v_n\dots i_r}_{j_1\dots j_s}
-\sum_{n=1}^s\sum_{w_n=1}^2\Gamma_{k j_n}^{w_n}F^{i_1\dots i_r}_{j_1\dots w_n\dots  j_s}+
m\varphi_k F^{i_1\dots i_r}_{j_1\dots j_s}.
$$
If the pseudotensorial field $F$ has type $(r,s)$ and weight $m$, then the
pseudotensorial field $\nabla F$ has type $(r, s+1)$ and weight $m.$

{\bf Step 7.} Pseudovectorial fields {\boldmath $\alpha$} and {\boldmath $\beta$} are collinear, hence exists the coefficient $N, $ it is the pseudoinvariant of  weight 2, such that:
$\beta=3N\alpha.$ Then 
$${\xi^i}=d^{ij}\nabla_jN,\qquad {M}=-\alpha_i\xi^i,\qquad {\gamma}=-\xi-2\Omega\alpha,\qquad {\Gamma}=-\frac{d_{ij}\nabla_{\xi}\xi^i\xi^j}{M}.
$$
Here {\boldmath $\xi$} -- pseudovectorial field of  weight 3;  ${M}$ --  pseudoinvariant of weight 4;  
{\boldmath $\gamma$} -- pseudovectorial field of  weight 3;
${\Gamma}$ -- pseudoinvariant of weight 4. (See paper \cite{Sharipov1}).

\subsection{Invariants}

 In this paper we are interest the case $M\ne 0$, that  named
{\it the first case of intermediate degeneration}. 
By definition it means that also $N\ne 0$ and {\boldmath $ \gamma$}$ \ne 0.$ The basic invariants are
\begin{equation}\label{inv1}
I_1=\frac{M}{N^2},\quad I_2=\frac{\Omega^2}{N},\quad I_3=\frac{\Gamma}{M}.
\end{equation}
By differentiating invariants $I_1$, $I_2$, $I_3$ along pseudovectorial fields  {\boldmath $\alpha$} and {\boldmath $\gamma$}
we get  invariants
\begin{equation}\label{inv2}
\aligned
I_4&=\frac{\nabla_{{\alpha}}I_1}{N},\qquad\quad I_5=\frac{\nabla_{ {\alpha}}I_2}{N},\qquad \quad I_6=\frac{\nabla_{ {\alpha}}I_3}{N},\\
I_7&=\frac{(\nabla_{ {\gamma}}I_1)^2}{N^3},\qquad I_8=\frac{(\nabla_{ {\gamma}}I_2)^2}{N^3},\qquad I_9=\frac{(\nabla_{ {\gamma}}I_3)^2}{N^3}.
\endaligned
\end{equation}
Repeating this procedure more and more times, we can form an infinite sequence of invariants, adding six ones in each step.
 
So, to calculate the invariants we have to compute:
\begin{enumerate}
\item pseudovectorial field {\boldmath $\alpha$} $=(B,\,-A)^T$ of weight 2, see formula (\ref{alpha});
\item pseudovectorial field {\boldmath $\gamma$} of weight 3, see formulas (\ref{gamma1}), (\ref{gamma2});
\item pseudoinvariant $F$ of  weight 1, see formula (\ref{F});
\item pseudoinvariant $M$ of weight 4, see formulas (\ref{M1}), (\ref{M2});
\item pseudoinvariant $N$ of weight 2, see formula (\ref{N});
\item pseudoinvariant $\Omega$ of  weight 1, see formulas (\ref{Omega1}), (\ref{Omega2});
\item pseudoinvariant $\Gamma$ of weight 4, see formula (\ref{Gamma}).
\end{enumerate}

\subsection{Correlation between the semiinvariants}

No doubt the main part of the pseudoinvariants have been known previously.

At the work  E.Cartan \cite{Cartan} have adopted the following notations:
$$P=-a_4,\quad Q=-a_3,\quad R=-a_2,\quad S=-a_1,\quad A=-L_1, \quad B=-L_2.$$
 
At the work R.Liouville \cite{Liouville} were presented the semiinvariants $\nu_5,$ $w_1$, $i_2$ and the quantity $R_1$ (see rewiew  in \cite{BordagBandle}).  Here is a link between these quantities and  pseudoinvariants $F$, $\Omega$, $N$ and  quantity $H$:
 $$
 F^5=\nu_5,\quad H=L_1(L_2)_x-L_2(L_1)_x+3R_1,\quad \Omega=-w_1-\frac{\nu_5 a_4}{L_1^3}-4\frac{(L_1)_x R_1}{L_1^3},\quad N=\frac{i_2}{3}.
 $$
 Pseudovectorial field {\boldmath $\gamma$}, pseudoinvariant $M$ and pseudoinvariant $\Gamma$
 first appeared in the papers \cite{Sharipov1}, \cite{Sharipov2}, \cite{Sharipov3}.
At the present paper we use the new notation in order to be able to compute the chais of invariants
(\ref{inv2}).

\section{The main problem}

The main problem is to describe the equivalence classes of equations (\ref{eq}) from the first case of intermediate degeneration with the conditions $I_1=const\ne 0$ and $I_2=0$ under the general point transformation (\ref{zam}).
So, we investigate equations (\ref{eq}) such that
\begin{equation}\label{cond}
{ \alpha}\ne 0,\quad F=0,\quad M\ne 0,\quad I_1=const\ne 0,\quad I_2=0.
\end{equation}
It is easy to see that two sequences of the invariants  (\ref{inv2}) become trivial ones: $I_4=I_7=\dots =0$ and $I_5=I_8=\dots =0$.

According to papers  \cite{Babich}, \cite{BordagBandle}, each equation  (\ref{eq}) that
satisfies the relations $F=0$ and $I_2=0$ can be transformed into the form
\begin{equation}\label{model}
y''=P(x,y).
\end{equation}
For  equations (\ref{model}) holds the following relations: 
$$
A=P_{0.2}\ne 0,\quad B=0,\quad M=\frac{2A^2_{0.1}}{5}-\frac{AA_{0.2}}{3}\ne 0, \quad N=-\frac{A_{0.1}}{3},\quad I_2=0.
$$
Let us calculate the invariant $I_1$:
\begin{equation}\label{A}
I_1=\frac {M}{N^2}=\frac {18}{5}-3\frac {AA_{0.2}}{A^2_{0.1}}=\frac {18}{5}-3C_1=const\ne 0,\quad
\frac {AA_{0.2}}{A^2_{0.1}}=C_1=const\ne \frac 65.
\end{equation}

\section{Four Types of equations}

\begin{theorem}\label{the1}  Every equation (\ref{eq}) with conditions (\ref{cond})
can be transformed by point transformations (\ref{zam}) into the form:
$$
y''=P^*(y)+t(x)y+s(x),
$$ 
where
$$
P^*(y)=\left\{ 
\aligned
& e^y,\quad \phantom{.....................}\mbox{if} \quad  I_1= \frac 35;\\
 -&\ln y, \quad \phantom{..................} \mbox{if} \quad I_1=-\frac {9}{10};\\
& y(\ln y-1), \quad \phantom{........}\mbox{if} \quad  I_1=-\frac {12}{5};\\
& \frac {y^{C+2}}{(C+1)(C+2)},\quad \mbox{if} \quad  I_1=\frac {3(C+5)}{5C},\; C=const\ne -5,\,-2,\,-1,\,0.
\endaligned\right.
$$
\end{theorem}

{\bf Proof.} 
Let us resolve the differential equation (\ref{A}) with respect to the function $A(x,y)$. There are two possibilities:
\begin{equation}\label{C1}
\aligned
& C_1=1,\qquad A(x,y)=b(x)\cdot e^{a(x)y},\\
& C_1\ne 1,\qquad A(x,y)=\left( {(a(x)y+b(x)}\right)^{C},\quad C=\frac 1{1-C_1}, 
\endaligned
\end{equation}
where $a(x)$ and $b(x)$ are arbitrary functions.

According to paper \cite{BordagBandle}, the most general  point transformations
preserving the form (\ref{model}) is the following transformation:
\begin{equation}\label{form}
x=\alpha\int p^2(\tilde x) d(\tilde x)+\beta,\quad y=p(\tilde x)\tilde y+h(\tilde x).
\end{equation}
Here $\alpha$, $\beta$ -- constants, $p(\tilde x)$, $h(\tilde x)$ -- certain functions.

Therefore the direct and inverse transformations matrices $S$ and $T$ are
$$
\aligned
S=&
\left(\begin{array}{cc}
	 \partial x /\partial \tilde  x& \partial x /\partial  \tilde y\\ \partial y /\partial  \tilde x & \partial  y /\partial \tilde y
	 \end{array}\right)=
\left(
\begin{array}{rr}
 \alpha p^2(\tilde x) & 0 \\ 
p_{1.0}(\tilde x)\tilde y+h_{0.1}(\tilde x)  & p(\tilde x)
 \end{array}
\right),\quad \det S=\alpha p^3(\tilde x),\\
 T=&S^{-1}=
\left(
\begin{array}{rr}
  \dfrac{1}{\alpha p^2(\tilde x)} & 0 \\ 
\dfrac{-p_{1.0}(\tilde x)\tilde y-h_{0.1}(\tilde x) }{\alpha p^3(\tilde x)} & \dfrac 1{ p(\tilde x)}
 \end{array}
\right),\qquad \qquad\quad\;\;\det T= \dfrac{1}{\alpha p^3(\tilde x)}.
\endaligned
$$
After the transformations (\ref{form}) the pseudovectorial field ${\alpha} $ of  weight 2 changes as
$$
\left(\begin{array}{r}
\tilde B \\ -\tilde A
\end{array}\right)= \frac{T}{(\det T)^2}\left(\begin{array}{r}
 0 \\ -A
\end{array}\right)=
\left(\begin{array}{r}
0 \\ -\alpha p^5(\tilde x) A
\end{array}\right).
$$
So the transformation rules for $A$ and $B$ are:
$
\tilde A(\tilde x,\tilde y)=\alpha^2p^5(\tilde x) A(x(\tilde x), y(\tilde x,\tilde y)),$ $ \tilde B=0.$

Let $C_1=1$ and function $A$ from (\ref{C1}), then
$$
\tilde A(\tilde x,\tilde y)=\alpha^2p^5(\tilde x) b(x(\tilde x))\cdot e^{a(x(\tilde x))(p(\tilde x)\tilde y+h(\tilde x))}=
\alpha^2p^5(\tilde x)b(x(\tilde x)) e^{a(x(\tilde x))h(\tilde x)} e^{a(x(\tilde x))p(\tilde x)\tilde y}.
$$
Choosing the appropriate functions $h(\tilde x)$ and $p(\tilde x)$ we can get 
$$
\alpha^2p^5(\tilde x)b(x(\tilde x)) e^{a(x(\tilde x))h(\tilde x)}=1,\quad a(x(\tilde x))p(\tilde x)=1,\quad \tilde A(\tilde x,\tilde y)=e^{\tilde y}.$$

Let $C_1\ne 1$ and function $A$ from (\ref{C1}), then
$$
\tilde A(\tilde x,\tilde y)=\alpha^2p^5(\tilde x)\left( {a(x(\tilde x))p(\tilde x)\tilde y+\left[a(x(\tilde x))h(\tilde x)+b(x(\tilde x)\right)]}\right)^{C}
$$
Choosing the appropriate functions $h(\tilde x)$ and $p(\tilde x)$ we can get 
$a(x(\tilde x))h(\tilde x)+b(x(\tilde x))=0,$ $
\alpha^2p^{5+C}(\tilde x)a^{C}(x(\tilde x))=1,$ $
\tilde A(\tilde x,\tilde y)=\tilde y^{C}.$

So, if equation (\ref{eq}) with conditions (\ref{cond}) is written in terms of
canonical coordinates (let these coordinates will be $(x,y)$), there may be two possibilities: $ A(x,y)=e^{ y}$ or $A(x,y)=y^{C}$.
Therefore $A=P_{0.2}$  then it may be 4 opportunities:
$$
\aligned
P(x,y)=& {e^{y}}+t(x)y+s(x),\quad \phantom{....................}C_1=1,\\
P(x,y)=& -{\ln y}+t(x)y+s(x),\quad \phantom{..............}C=-2,\\
P(x,y)=& y\ln y-y+t(x)y+s(x),\quad \phantom{..........} C=-1,\\
P(x,y)=& \frac{y^{C+2}}{ (C+1)(C+2)}+t(x)y+s(x),\quad C=const\ne -5,\,-2,\,-1,\,0.
\endaligned
$$
Here $t(x)$, $s(x)$ are the {arbitrary functions}.

\begin{table}[ht]
{Four different types of  equations (\ref{eq}) with conditions (\ref{cond}).}
{\begin{tabular}{@{}clcccc@{}} \hline
 Type & Equation & $A$ & $C$ & $C_1$ & $I_1$ \\
\hline
I & $y''=e^y+t(x)y+s(x)$ & $e^y$ & - & 1 & $\frac 35$ \\
II & $y''=-\ln y+t(x)y+s(x)$ & $\frac 1{y^2}$ & -2 & $\frac 32$ & $-\frac 9{10}$ \\
III & $y''=y(\ln y-1)+t(x)y+s(x)$ & $\frac 1y$ & -1 & 2 & $-\frac{12}{5}$ \\
IV & $y''=\frac{y^{C+2}}{(C+1)(C+2)}+t(x)y+s(x)$ & $y^C$ & $C\ne 0,-1, -2,-5$ & $\frac {C-1}C$ & $ \frac {3(C+5)}{5C}$\\
\hline
\end{tabular}}
\end{table}

\section{  Equations of Type I }

\begin{definition}
Let us say that equation (\ref{eq}) has Type I if  conditions (\ref{cond}) hold, where  $I_1={3}/{5}$. 
\end{definition}

According to Theorem \ref{the1} any equation (\ref{eq}) of Type I
can be transformed by point transformations (\ref{zam}) into the canonical form:
\begin{equation}\label{typeI}
y''=e^{y}+t(x)y+s(x).
\end{equation}

\begin{lemma}\label{le1} The most general point transformations that preserve the canonical form (\ref{typeI}) are the following ones:
\begin{equation}\label{trType1}
x=\alpha \tilde x+\beta,\quad y=\tilde y-2\ln\alpha.
\end{equation}
Here $\alpha$, $\beta$ are some constants.
The new equation has the form: $ \tilde y''=e^{\tilde y}+\tilde t (\tilde x)\tilde y+\tilde s(\tilde x),$
where
$$
\tilde t (\tilde x)=\alpha^2 t(\alpha \tilde x+\beta),\quad
\tilde s (\tilde x)=\alpha^2 s(\alpha \tilde x+\beta)-2 \alpha^2 \ln\alpha \cdot t(\alpha \tilde x+\beta).
$$
\end{lemma}

The proof of Lemma \ref{le1} follows from the straightward calculations. We apply transformations (\ref{form}) to the equation 
(\ref{typeI}). They must preserve the form of equation.
Then in these canonical coordinates the non-trivial invarians (\ref{inv1}), (\ref{inv2}), (\ref{inv6921})  are equal:
$$
\aligned
I_3& =\frac{1}{15}+\frac{1}{15e^{y}}\left(t(x)y+s(x)\right),\quad I_6=\frac{1}{5e^{y}}\left(t(x)y+s(x)-t(x)\right),\\
I_9&=\frac{1}{1875 e^{3y}}\left(t'(x)y+s'(x)\right)^2.
\endaligned
$$

Let us introduce the additional invariants:
\begin{equation}\label{invT1}
\aligned
J_3&=15I_3-1=\frac{t(x)y+s(x)}{e^{y}},\quad\quad J_6=5I_6=\frac{t(x)y+s(x)-t(x)}{e^{y}},\\
J&=\frac{J_3}{J_3-J_6}=y+\frac{s(x)}{t(x)},\quad\qquad\quad J_1=J+\ln\left(\frac{J_3}{J}\right)=\ln t(x)+\frac{s(x)}{t(x)},\\
J_9& =1875I_9=\frac{\left(t'(x)y+s'(x)\right)^2}{e^{3y}},\quad K=\frac{J_9}{J_3^3}=\frac{\left(t'(x)y+s'(x)\right)^2}
{\left(t(x)y+s(x)\right)^3}.
\endaligned
\end{equation}

\vskip 0.5cm

\begin{theorem}\label{the2}  
Let equation (\ref{eq}) be an arbitrary equation of Type I. Then it is equivalent to some equation from the following list of nonequivalent equations of Type I:
\begin{enumerate}
 \item If $J_3=0$ from (\ref{invT1}) then it is equivalent to  $y''=e^{y}.$
 \item If $J_3\ne const$, $J_3=J_6$, $K=0$ from (\ref{invT1}) then it is equaivalent to  $y''=e^{y}+1.$
 \item If $J_3\ne const$, $J_3\ne J_6$, $J_6\ne const$, $J_1=a=const$, $K=0$ from (\ref{invT1}) then it is equaivalent to  $y''=e^{y}+y+a.$

\noindent Two equations of Type I.3 are equivalent if and only if  invariant $J_1$ for both equations is the same and equal to constant $a$.
 \item If $J_3\ne const$, $J_3=J_6$, $K=k=const\ne 0$ from (\ref{invT1}) then it is equaivalent to  $y''=e^{y}+{4}/{kx^2}.$

\noindent Two equations of Type I.4 are equivalent if and only if  invariant $K$ for both equations is the same and equal to constant $k\ne 0$.
\item If $J_3\ne const$, $J_3=J_6$, $K\ne const$ from (\ref{invT1}) then it is equaivalent to  $y''=e^{y}+s(x), $ $ s(x)\ne const.$

\noindent Two equations of Type I.5 are equivalent if and only if after the transformation $\tilde x=K(x,y),\, \tilde y=J_3(x,y)$ their
notations become identical.
 \item If $J_3\ne const$, $J_3\ne J_6$, $J_6\ne const$, $J_1\ne const$, $K\ne const$ from (\ref{invT1}) then it is equaivalent to  $y''=e^{y}+t(x)y+s(x),\quad t(x)\ne 0.$

\noindent Two equations of Type I.6 are equivalent if and only if after the transformation $\tilde x=J_1(x,y),\, \tilde y=J(x,y)$ their
notations become identical.

\end{enumerate}
In the cases I.5 and I.6 functions $t(x)$ and $s(x)$ are definited up to
transformations (\ref{trType1}).
\begin{table}[ht]
{Equations of Type I.}
{\begin{tabular}{@{}cccccl@{}} \hline
Type & $J_3$ & $J_6$ & $J_1$ & $K$ & Canonical form\\
 \hline
I.1 &  $0$ &  $0$ &   $-$  &  $0$ &   $y''=e^{y} $\\ 
I.2 & $\ne$ const & $J_3$ &   $-$  &  $0$ &  $y''=e^{y} +1$\\ 
I.3 & $\ne$ const &  $\ne J_3,$ &   const=a  & $0$ &  $y''=e^{y} +y+a$\\  
      &  &           $ \ne$ const &  & & \\
I.4 & $\ne$ const &  $J_3$ & $-$   & $const=k\ne 0$ & $y''=e^{y}+\frac{4}{kx^2}$\\ 
I.5 & $\ne$ const &  $J_3$ &    $-$ &  $\ne$ const  & $y''=e^{y} +s(x),$ $s(x)\ne$  const\\
I.6 & $\ne$ const &  $\ne J_3,$  &   $\ne $ const  &  $\ne $ const & $y''=e^{y} +t(x)y+s(x),$ \\ 
    & &             $ \ne$ const &   &  & $t(x)\ne 0$  \\
\hline
\end{tabular}}
\end{table}
\end{theorem}

{\bf Proof.} Let  equation (\ref{eq}) be an arbitrary equation of Type I. Then in the terms of canonical coordinates it has the form (\ref{typeI}). Let us calculate the invariants (\ref{invT1}).
\begin{enumerate}
 \item  If $J_3=0$ then 
$({t(x)y+s(x)})/{e^{y}}=0$ $ \Longleftrightarrow $ $t(x)\equiv 0,$ $ s(x)\equiv 0.$
Therefore  equation (\ref{eq}) can be reduced into the canonical form $y''=e^{y}. $

\item If $J_3\ne const$, $J_3=J_6$, $K=0$, then 
$$
\frac{t(x)y+s(x)}{e^{y}}=\frac{t(x)y+s(x)-t(x)}{e^{y}},\quad \frac {\left(t'(x)y+s'(x)\right)^2}
{\left(t(x)y+s(x)\right)^3}=0. $$
Hence $t(x)\equiv 0,$  $s'(x)\equiv 0$, accordingly $s(x)=s=const\ne 0.$
If $s=0$ then $J_3=0$ and equation has Type I.1.
So in the terms of canonical coordinates this equation has the form
$y''=e^{y}+s.$
Let us make the transformation (\ref{trType1}) then  
$\tilde s=\alpha^2 s.$ Choosing the parameter $\alpha$ we could make $\tilde s=1.$ Thus the canonical form is $\tilde y''=e^{\tilde y}+1.$

\item If $J_3\ne const$, $J_3\ne J_6$, $J_6\ne const$, $J_1=a=const$, $K=0$ then
$$ \ln t(x)+\frac{s(x)}{t(x)}=a,\quad \frac{\left(t'(x)y+s'(x)\right)^2}
{\left(t(x)y+s(x)\right)^3}=0. $$
Hence $t(x)=t=const\ne 0,$ $s(x)=s=const$ and equation has the form
$y''=e^{y}+ty+s.$
Let $Y''=e^{Y}+TY+S,$  $ T=const\ne 0,$ $S=const$ is another equation of this type. 
These equations are equivalent  if and only if 
$
T=\alpha^2t,$ $ S=\alpha^2s-\alpha^2t\ln \alpha^2$
then between $t,\,s$ and $T,\,S$   exists the following connection
$$
S=\frac{sT}{t}-T\ln\left(\frac Tt\right)\;\; \Leftrightarrow\;\; \ln t+\frac{s}{t}=\ln T+\frac{S}{T}\;\;\Leftrightarrow\;\; 
J_1(x,y)=J_1(X,Y)=a.
$$
Thus we see that two equations of Type I.3 are equivalent  if and only if  invariants $J_1$ for both equations are equal
to constant $a$.

Now let us select the canonical form for equations of Type I.3.
We make transformation (\ref{trType1}) and choose the parameter $\alpha$ so that into the new coordinates
$\tilde t=1,$ then invariant $J_1=\tilde s=a.$
Hence equation has the canonical form
$\tilde y''=e^{\tilde y}+\tilde y+a.$

\item If $J_3\ne const$, $J_3=J_6$, $K=k=const\ne 0$, then
$$
\frac{t(x)y+s(x)}{e^{y}}=\frac{t(x)y+s(x)-t(x)}{e^{y}},\quad \frac{\left(t'(x)y+s'(x)\right)^2}
{\left(t(x)y+s(x)\right)^3}=k.
$$
Hence $t(x)\equiv 0$ and we have a differential equation on function $s(x)$: $s^{\prime 2}(x)=ks^3(x)$.
  The solution:
$s(x)={4}/{(\sqrt k\cdot x+c)^2},$   $ c=const.$
Now let us choose the canonical form for the equations of Type 1.4.
Making the transformation (\ref{trType1}) we get
$$
\tilde s(\tilde x)=\alpha^2\cdot s(\alpha \tilde x+\beta)=\frac{4\alpha^2}{(\alpha\sqrt k\cdot \tilde x+\sqrt k\cdot\beta+c)^2},
$$
then select the parameter  $\beta$ such that $\sqrt k\cdot\beta+c=0$ for any $\alpha$. So  equation has the canonical form
$
\tilde y''=e^{\tilde y}+{4}/{(k\tilde x^2)}.
$
Two equations of Type I.4 equivalent if and only if  invariants $K$ for both equations are equal to constant $k\ne 0$.

\item If $J_3\ne const$, $J_3=J_6$, $K\ne const$ then 
$$
\frac{t(x)y+s(x)}{e^{y}}=\frac{t(x)y+s(x)-t(x)}{e^{y}},\quad\mbox{so}\quad t(x)\equiv 0.
$$
Then in terms of canonical coordinates  equation has the form
$y''=e^{y}+s(x).$

Let us solve the equivalence problem for the equations of Type I.5.
We have two possibilities. One way is to reduce both equations into the canonical coordinates:
$y''=e^{y}+s(x),$ $ Y''=e^{Y}+S(X).$
These equations are equivalent if and only if there exist appropriate $\alpha$ and $\beta$ such that
$S (X)=\alpha^2 s(\alpha X+\beta).$

The other way is based on the observation that for any equation (\ref{eq}) of Type I.5 invariants $K$ and $J_3$ are functionally independent.  
So, if the first equation has the form (\ref{eq}) and depends on  coordinates $(x,y)$ and the second equation 
has the form (\ref{eq}) and depends on  coordinates $(X,Y)$, 
 we can  make the invariant point transformation
$$
\aligned
\tilde x&=K(x,y),\quad \tilde y=J_3(x,y)\quad\mbox{for the first equation,} \\
\tilde x&=K(X,Y),\;\; \tilde y=J_3(X,Y)\quad\mbox{for the second equation.}
\endaligned
$$
Equations are equivalent if and only if in the term of coordinates $(\tilde x,\, \tilde y)$ their
notations become identical.

 \item If $J_3\ne const$, $J_3\ne J_6$, $J_6\ne const$, $J_1\ne const$, $K\ne const$ then  
 in terms of canonical coordinates equation has form  
$y''=e^{y}+t(x)y+s(x),\quad t(x)\ne 0.$

Let us solve the equivalence problem for the equations of Type I.6. 
As at the previous case we have two possibilities.

The first way: in terms of canonical coordinates these equations have forms
$$
y''=e^{y}+t(x)y+s(x),\quad Y''=e^{Y}+T(X)Y+S(X). 
$$
They equivalent if and only if exist constants $\alpha$ and $\beta$ such that
$$
T (X)=\alpha^2 t(\alpha X+\beta),\quad
S (X)=\alpha^2 s(\alpha X+\beta)-2 \alpha^2 \ln\alpha \cdot t(\alpha X+\beta).
$$

The second way:
note that for any equation (\ref{eq}) of Type I.6  invariants $J_1$ and $J$ are functionally independent. 
So, 
we can  make the invariant point transformation
$$
\aligned
\tilde x&=J_1(x,y),\quad  \tilde y=J(x,y)\quad\mbox{for the first equation,} \\
\tilde x&=J_1(X,Y),\;\; \tilde y=J(X,Y)\quad\mbox{for the second equation.}
\endaligned
$$
Equations are equivalent if and only if in the term of coordinates $(\tilde x,\tilde y)$ their
notations become identical.
\end{enumerate}

\section{  Equations of Type II }

\begin{definition}
Let us say that equation (\ref{eq}) has Type II if conditions (\ref{cond}) hold, where  $I_1=-{9}/{10}$. 
\end{definition}

According to Theorem \ref{the1} any equation (\ref{eq}) of Type II
can be reduced by point transformations (\ref{zam}) into the canonical form:
\begin{equation}\label{typeII}
y''=-\ln {y}+t(x)y+s(x).
\end{equation}

\begin{lemma}\label{le2} The most general point transformations that preserve the canonical form (\ref{typeII}) are the following ones:
\begin{equation}\label{trTypeII}
 x=\alpha^{-\frac 13} \tilde x+\beta,\quad y=\alpha^{-\frac 23}\tilde y.
\end{equation}
Here $\alpha$, $\beta$ are some constants.
In the new coordinates this equation has the following form: $\tilde y''=-\ln {\tilde y}+\tilde t (\tilde x)\tilde y+\tilde s(\tilde x),$
where
$$
\tilde t (\tilde x)=\alpha^{-\frac 23} t(\alpha^{-\frac 13} \tilde x+\beta),\quad
\tilde s (\tilde x)=s(\alpha^{-\frac 13} \tilde x+\beta)+\frac 23 \ln \alpha.
$$
\end{lemma}


Then in terms of canonical coordinates the non-trivial invarians (\ref{inv1}), (\ref{inv2}), (\ref{inv6921}):
$$
I_3=\frac{2}{5}\ln y-\frac{2}{5}\left(t(x)y+s(x)\right),\quad I_6=\frac 35-\frac{3}{5}y t(x),\quad
I_9=-\frac{54}{625}y\left(t'(x)y+s'(x)\right)^2.
$$

Let us introduce the additional invariants
\begin{equation}\label{invT2}
 \aligned
J_3&=\frac{5I_3}{2}=\ln y-t(x)y-s(x),\quad \quad J_6=\frac{3-5I_6}{3}=t(x)y,\\
J_9&=-\frac{625I_9}{54}=y\left(t'(x)y+s'(x)\right)^2,\quad J=\ln(J_6)-J_3-J_6=s(x)+\ln t(x).
\endaligned
\end{equation}

\vskip 0.5cm

\begin{theorem}\label{the3} 
Let equation (\ref{eq}) be an arbitrary equation of Type II. Then it is equivalent to some equation from the following list of nonequivalent equations of Type II:
\begin{enumerate}
 \item If $J_6=0$, $J_9=0$ from (\ref{invT2}) then it is  equaivalent to  $y''=-\ln y$.
\item If $J_6\ne const$, $J_9=0$, $J=a$ from (\ref{invT2}) then it is equaivalent to  
$y''=-\ln y+y+a$.

\noindent Two equations of Type II.2 are equivalent if and only if invariant $J$ for both equations is the same and equals to constant $a$.
 \item If $J_6=0$,  $J_9\ne const$ from (\ref{invT2}),  then the equation is equaivalent to  $y''=-\ln y+s(x)$, $s(x)\ne const$.

\noindent Two equations of Type II.3 are equivalent if and only if after the transformation $\tilde x=J_3(x,y),\, \tilde y=J_9(x,y)$ their
notations become identical.
	\item If $J_6\ne const$,  $J_9\ne const$ from (\ref{invT2})  then it is equivalent to  $y''=-\ln y+t(x)y+s(x)$, $t(x)\ne 0$.

\noindent Two equations of Type II.4 are equivalent if and only if after the transformation $\tilde x=J(x,y),\, \tilde y=J_6(x,y)$ their
notations become identical.
\end{enumerate}
In the cases II.2, II.4  functions $t(x)$, $s(x)$ are definited up to transformations (\ref{trTypeII}).
\begin{table}[ht]
{Equations of Type II.}
{\begin{tabular}{@{}ccccl@{}} \hline
Type &  $J_6$ & $J_9$ & $J$ & Canonical form\\
 \hline
II.1 &  $0$ &  $0$ &   $-$  &    $y''=-\ln {y} $\\ 
II.2 & $\ne$ const & 0 &   $a=$const$\ne 0$  &   $y''=-\ln {y}+y+a$\\ 
II.3 & 0 &  $\ne$ const, &   -  &  $y''=-\ln {y}+s(x),$ $s(x)\ne const$\\  
II.4 & $\ne$ const &  $\ne$const  & $\ne$const & $y''=-\ln {y}+t(x)y+s(x),$ $t(x)\ne 0$\\
\hline
\end{tabular}}
\end{table}
\end{theorem}

{\bf Proof.}  
Let us have the certain equation (\ref{eq}) of Type II. Then in terms of canonical coordinates it has form (\ref{typeII}).

\begin{enumerate}
 \item If $J_6=0$ and $J_9=0$ then 
$t(x)y=0,$ $ y\left(t'(x)y+s'(x)\right)^2=0$ $ \Longleftrightarrow $ $t(x)\equiv 0,$ $s(x)=s=const $
and equation may be reduced into the  form $y''=-\ln y+s.$
Let us make transformations (\ref{trTypeII}). Then $\tilde s=s+ 2/3 \ln \alpha$.
Choosing the appropriate $\alpha$ we can make $\tilde s=0$ then equation will be
$\tilde y''=-\ln \tilde y. $
\item If $J_6\ne const$, $J_9=0$, $J=a$, then
$
y\left(t'(x)y+s'(x)\right)^2=0$ $\Longleftrightarrow $ $ t'(x)\equiv 0,$ $ s'(x)\equiv 0.$ 
Accordingly $t(x)=t=const,$ $s(x)=s=const$ and equation will be
$y''=-\ln y+ty+s,$ $ t\ne 0.$
Let we have another equation of Type II.2  such that in terms of canonical coordinates
it has the form
$Y''=-\ln Y+TY+S,$ $T=const,$ $ T\ne 0,$   $ S=const.$
These equations equivalent if and only if
$$
T=\alpha^{-\frac 23} t,\quad S=s+\frac 23 \ln \alpha\;\; \Leftrightarrow\;\; 
S+\ln T=s+\ln t\;\; \Leftrightarrow\;\; J(x,y)=J(X;Y)=a.
$$
So we see that two equations of Type II.2 are equivalent if and only if invariants $J$ for both equations are equal to constant $a$.

Now let us find the canonical form for the equations of Type II.2. After the  transformations (\ref{trTypeII}) 
$\tilde t=\alpha^{-\frac 23} t,$ $\tilde s =s+ 2/3 \ln \alpha.$
Choosing the appropriate $\alpha$ we can get $\tilde t=1.$ 
Consequently the canonical form is
$\tilde y''=-\ln\tilde y+\tilde y+a.$

\item  If $J_6=0$,  $J_9\ne const$, then $t(x)\equiv 0,$ $s'(x)\ne 0$. Therefore in terms of canonical coordinates
equation has form
$y''=-\ln y+s(x),$ $s(x)\ne const.$
 
As at the previous case for the equations of Type I we have two possibilities.
The first way: in terms of canonical coordinates two equations have forms
$$
y''=-\ln y+s(x),\quad Y''=-\ln Y+S(X).
$$
They equivalent if and only if exist constants $\alpha$ and $\beta$ such that
$S(X)=s(\alpha^{-\frac 13} X+\beta)+ 2 \ln \alpha/3. $
The second way:
note that for any equation (\ref{eq}) of the Type II.3 invariants $J_3$ and $J_9$ are functionally independent. 
So, 
 we can  make the invariant point transformation
$$
\aligned
\tilde x&=J_3(x,y),\; \tilde y=J_9(x,y)\quad\mbox{for the first equation,} \\
\tilde x&=J_3(X,Y),\; \tilde y=J_9(X,Y)\quad\mbox{for the second equation.}
\endaligned
$$
These equations are equivalent if and only if in terms of new coordinates $(\tilde x,\tilde y)$ their
notations become identical.

\item If $J_6\ne const$,  $J_9\ne const$  then equation is equivalent to  
$y''=-\ln y+t(x)y+s(x),$ $t(x)\ne 0.$ 
The first way: in terms of canonical coordinates two equations have forms
$$
y''=e^{y}+t(x)y+s(x),\quad Y''=e^{Y}+T(X)Y+S(X).
$$
They equivalent if and only if exist the constants $\alpha$ and $\beta$ such that
$$
T(X)=\alpha^{-\frac 23} t(\alpha^{-\frac 13} X+\beta),\quad
S (X)=s(\alpha^{-\frac 13} X+\beta)+\frac 23 \ln \alpha.
$$

The second way:
note that for any equation (\ref{eq}) of the Type II.4 invariants $J$ and $J_6$ are functionally independent. 
So, 
 we can  make the invariant point transformation
$$
\tilde x=J(x,y),\; \tilde y=J_6(x,y),\quad \tilde  x=J(X,Y),\; \tilde y=J_6(X,Y)
$$
These equations are equivalent if and only if in terms of new coordinates $(\tilde x,\tilde y)$ their
notations become identical.
\end{enumerate}

\section{  Equations of Type III }
\begin{definition}
Let us say that equation (\ref{eq}) has Type III if conditions (\ref{cond}) hold, where  $I_1=-{12}/{5}$.
\end{definition}

According to Theorem \ref{the1} any equation (\ref{eq}) of the Type III
can be reduced by point transformations (\ref{zam}) into the canonical form:
\begin{equation}\label{typeIII}
y''=y(\ln {y}-1)+t(x)y+s(x).
\end{equation}

\begin{lemma}\label{le3}  The most general point transformations that preserve the canonical form (\ref{typeIII}) are the following ones:
\begin{equation}\label{trTypeIII}
 x= \pm\tilde x+\beta,\quad y=\frac{\tilde y}{\sqrt{\alpha}}.
\end{equation}

Here $\alpha$, $\beta$ are certain constants.
In term of new coordinates equation has form:
$ \tilde y''=\tilde y(\ln {\tilde y}-1)+\tilde t (\tilde x)\tilde y+\tilde s(\tilde x),$
where
$$
\tilde t (\tilde x)= t(\pm\tilde x+\beta)-\frac{\ln\alpha}{2},\quad
\tilde s (\tilde x)= \sqrt{\alpha} \cdot s(\pm\tilde x+\beta).
$$
\end{lemma}

Then in term of canonical coordinates  the non-trivial invarians (\ref{inv1}), (\ref{inv2}), (\ref{inv6921}):
$$
I_3=\frac{4}{15}(1-\ln y)-\frac{4}{15y}\left(t(x)y+s(x)\right),\;I_6=-\frac{4}{5}-\frac{4 s(x)}{5y},\;
I_9=-\frac{256}{1875y^2}\left(t'(x)y+s'(x)\right)^2.
$$

Let us denote the additional invariants
\begin{equation}\label{invT3}
\aligned
J_3&=\frac{4-15I_3}{4}=\ln y+t(x)+\frac{s(x)}{y},\quad J_6=\frac{1+5I_6}{4}=\frac{s(x)}{y},\\
J& =J_6e^{J_3-J_6}=s(x)e^{t(x)},\quad J_9=-\frac{1875I_9}{256}=\frac{\left(t'(x)y+s'(x)\right)^2}{y^2}.
\endaligned
\end{equation}

\begin{theorem}\label{the4}  
Let equation (\ref{eq}) be an arbitrary equation of Type III. Then it is equivalent to some equation from the following list of nonequivalent equations of Type III:
\begin{enumerate}
 \item If $J_6=0$, $J_9=0$ from (\ref{invT3}) then it is equaivalent to  $y''=y(\ln y-1)$.
\item If $J_6=0$, $J_9=b^2=const\ne 0$ from (\ref{invT3}) then it is equaivalent to  $y''=y(\ln y-1)\pm bxy$.

\noindent Two equations of Type III.2 are equivalent if and only if  invariant $J_9$ (\ref{invT3}) for both equations is the same and equals to the constant $b^2$.
\item If $J_6\ne const$, $J_9=0$, $J=a$ from (\ref{invT3})  then it is equaivalent to  $y''=y(\ln y-1)+a$.

\noindent Two equations of Type III.3 are equivalent if and only if  invariant $J$ (\ref{invT3}) for both equations is the same and equals to constant $a$.

 \item If $J_6\ne const$,  $J_9=b^2=const\ne 0$ from (\ref{invT3})  then it is equaivalent to  $y''=y(\ln y-1) \pm bxy+1$.

\noindent Two equations of Type III.4 are equivalent if and only if  invariant $J_9$ (\ref{invT3}) for both equations is the same and equals to constant $b^2$.
   \item If $J_6\ne const$,  $J_9\ne const$ from (\ref{invT3})   then it is equaivalent to  $y''=y(\ln y-1)+t(x)y+s(x)$, $s(x)\ne 0$.

\noindent Two equations of Type III.4 are equivalent if and only if after the transformation $\tilde x=J(x,y),\, \tilde y=J_6(x,y)$ their notations become identical.
\end{enumerate}
In the case III.5 functions $t(x)$ and $s(x)$ are definited up to transformations (\ref{trTypeIII}).
\begin{table}[ht]
{Equations of Type III.}
{\begin{tabular}{@{}ccccl@{}} \hline
Type &  $J_6$ & $J_9$ & $J$ & Canonical form\\
 \hline
III.1 &  $0$ &  $0$ &   $0$  &    $y''=y(\ln y-1) $\\ 
III.2 & $0$  & $b^2=$const$\ne 0$  &   0 &   $y''=y(\ln y-1)\pm bxy$\\ 
III.3 & $\ne$ const &  0 &   $a=$const$\ne 0$  &  $y''=y(\ln y-1)+a$\\
III.4 & $\ne$ const &  $b^2=$const$\ne 0$  &   $\ne$ const  &  $y''=y(\ln y-1)\pm bxy+1$\\
III.5 & $\ne$ const &  $\ne$ const  & $\ne$ const & $y''=y(\ln y-1)+t(x)y+s(x),$\\
 & & &    & $\phantom{.....}s(x)\ne 0$\\
\hline
\end{tabular}}
\end{table}
\end{theorem}

{\bf Proof.} 
Let us have the certain equation (\ref{eq}) of Type III. Then in terms of canonical coordinates it has the form (\ref{typeIII}).

\begin{enumerate}
 \item If $J_6=0$ and $J_9=0$, then $s(x)\equiv 0$, $t'(x)\equiv 0.$ 
Therefore $t(x)=t=const.$
Choosing the parameter $\alpha$ from the point transformation (\ref{trTypeIII}) we can 
get $\tilde t=t-{\ln\alpha}/{2}=0.$ Thus the canonical form is
$ \tilde y''=\tilde y(\ln {\tilde y}-1).$

\item If $J_6=0$ and $J_9=b^2=const$ then $s(x)\equiv 0,$ $t'(x)=\pm b=const$. So $t(x)=\pm bx+t,$ $t=const.$
Let's make the point transformation (\ref{trTypeIII}):
$\tilde t(\tilde x)= \pm b(\pm\tilde x+\beta)+t-{\ln\alpha}/{2}=\pm b\tilde x+(\pm b\beta+t-{\ln\alpha}/{2}).$
Choosing the parameters $\alpha$ and $\beta$ we get $\tilde t(\tilde x)=\pm b\tilde x.$

It is easy to check that two equation of Type III.2 are equivalent if and only if 
  invariants $J_9$ (\ref{invT3}) for both equations are equal to the constant $b^{2}$.

\item If $J_6\ne const$, $J_9=0$,  then $t'(x)\equiv $ and $s'(x)\equiv 0,$ $s(x)\ne 0.$
Hence $t(x)=t=const,$ $s(x)=s=const\ne 0.$  
Let's make the point transformation (\ref{trTypeIII}):
$\tilde t = t-{\ln\alpha}/{2},$ $\tilde s = \sqrt{\alpha} \cdot s.$
Choosing the parameter $\alpha$ we can make $\tilde t=0.$ 
At the new coordinates $J=\tilde s=a.$
Thus two equations of Type III.3 are equivalent if and only if  invariants $J$ for both equations are equal to the constant $a.$

\item If $J_9=const\ne 0$ then $s'(x)\equiv 0$ and $t'(x)=\pm b=const \ne 0.$ So $t(x)=\pm bx+t,$ $t=const,$ $s(x)=s=const\ne 0.$
Hence
$
\tilde t (\tilde x)= \pm b(\pm\tilde x+\beta)+t-{\ln\alpha}/{2},$ $
\tilde s = \sqrt{\alpha} \cdot s.
$
Choosing the parameters $\alpha$ and $\beta$ we can make $\tilde t=\pm b\tilde x,$ $\tilde s=1.$ 
So in terms of new coordinates equation has form $
\tilde y''=\tilde y(\ln \tilde y-1) \pm b\tilde x\tilde y+1.$
It is easy to check that two equation of Type III.4 are equivalent if and only if 
 invariants $J_9$ (\ref{invT3}) for both equations are equal to the constant $b^{2}$. 

\item If $J_3\ne const,$ $J_9\ne const$ then  equation has form
$y''=y(\ln y-1)+t(x)y+s(x),$ $s(x)\ne 0.$
The first way:  two equations of Type III.5 are equivalent if and only if in terms of canonical coordinates
 condition $ t(\pm X+\beta)+\ln s(\pm X+\beta)=T(X)+\ln S (X)$ holds, 
where the second equation in terms of canonical coordinates has form

\noindent $ Y''=Y(\ln Y-1)+T(X)Y+S(X),$ $S(X)\ne 0.$

The second way: note that for any  equation (\ref{eq}) of Type III.5  invariants $J$ and $J_6$ are functionally independent. 
Hence we can make the invariant point transformation:
$$
\tilde x=J(x,y),\; \tilde y=J_6(x,y),\quad \tilde x=J(X,Y),\; \tilde y=J_6(X,Y).
$$
Equations are equivalent if and only if in terms of coordinates $(\tilde x,\tilde y)$ their notations become identical.
\end{enumerate}

\section{  Equations of Type IV }

\begin{definition}
Let us say that equation (\ref{eq}) has Type IV if conditions (\ref{cond}) hold, where  $I_1={3(C+5)}/{5C},$ $C=const$, $C \ne 0,\,-1,\,-2,\,-5$.
\end{definition}

According to Theorem \ref{the1} any equation (\ref{eq}) of Type IV
can be reduced by point transformations (\ref{zam}) into the canonical form:
\begin{equation}\label{typeIV}
y''=\frac{y^{C+2}}{(C+1)(C+2)}+t(x)y+s(x).
\end{equation}

\begin{lemma}\label{le4}  The most general point transformations that preserve the canonical form (\ref{typeIV}) are the following ones:
\begin{equation}\label{trTypeIV}
 x= \alpha^{\frac{C+1}{C+5}} \tilde x+\beta,\quad y=\alpha^{\frac{-2C}{C+5}}{\tilde y}.
\end{equation}

Here $\alpha$, $\beta$ are some constants.
In terms of new coordinates  equation has the form:
$
\tilde y''={\tilde y^{C+2}}/{((C+1)(C+2))}+\tilde t (\tilde x)\tilde y+\tilde s(\tilde x),
$
where
$$
\tilde t (\tilde x)= \alpha^{\frac{2(C+1)}{C+5}}  \cdot t(\alpha^{\frac{C+1}{C+5}} \tilde x+\beta),\quad
\tilde s (\tilde x)= \alpha^{\frac{2(C+2)}{C+5}}\cdot s(\alpha^{\frac{C+1}{C+5}} \tilde x+\beta).
$$
\end{lemma}

The basic invariants (\ref{inv1}), (\ref{inv2}), (\ref{inv6921}) are
$$
\aligned
&I_3=\frac {C(C+5)}{15(C+1)(C+2)}\cdot \frac{y^{C+2}+(t(x)y+s(x))(C+1)(C+2)}{y^{C+2}},\\
&I_6=\frac{(C+5)}5\cdot\frac{t(x)y(C+1)+s(x)(C+2)}{y^{C+2}},\\
&I_9=\frac{ C (C+5)^4 }{1875}\cdot\frac{\left(t'(x)y+s'(x)\right)^2 }{y^{3C+5}},\\
&I_{21}=\frac{(\nabla_{{\gamma}}I_9)^2}{N^3}=\frac{4C(C+5)^{10}}{29296875}\cdot\frac{\left(t'(x)y+s'(x)\right)^2\left(t''(x)y+s''(x)\right)^2}{y^{7C+11}}.
\endaligned
$$

The additional invariants 
\begin{equation}\label{invT4}
\aligned
&J_3=\frac{15(C+1)(C+2)I_3-C(C+5)}{C(C+1)(C+2)(C+5)}=\frac{t(x)y+s(x)}{y^{C+2}},\\
&J_6=\frac{5I_6}{C+5}=\frac{t(x)y(C+1)+s(x)(C+2)}{y^{C+2}},\quad J_9=\frac{1875}{C (C+5)^4 }=\frac{\left(t'(x)y+s'(x)\right)^2 }{y^{3C+5}},\\
&J_{21} =\frac{29296875I_{21}}{4C(C+5)^{10}}=\frac{\left(t'(x)y+s'(x)\right)^2\left(t''(x)y+s''(x)\right)^2}{y^{7C+11}},\\
&J_1=(C+2)J_3-J_6=\frac{t(x)}{y^{C+1}},\quad J_2=J_6-(C+1)J_3=\frac{s(x)}{y^{C+2}},\\
&J =\frac{J_2}{J_1^{\frac{C+2}{C+1}}}=\frac{s(x)}{t(x)^{\frac{C+2}{C+1}}}, \quad K=\frac{J_9}{J_1^3}=\frac{(t^{\prime }(x)y+s'(x))^2}{y^2t^3(x)},\\
&K_1=\sqrt{\frac{J_{21}}{J_9}}=\frac{t''(x)y+s''(x)}{y^{2C+3}},\quad K_2=\frac{K_1}{J_1^2}=\frac{t''(x)y+s''(x)}{yt^2(x)}.
\endaligned
\end{equation}

\vskip 0.5cm

\begin{theorem}\label{the5}  
Let equation (\ref{eq}) be an arbitrary equation of Type IV. Then it is equivalent to some equation from the following list of nonequivalent equations of Type  IV:
\begin{enumerate}
 \item If $J_1=0$,  $J_2=0$ from (\ref{invT4}) then it is equaivalent to  $y''=\frac{y^{C+2}}{(C+1)(C+2)}.$

\item If $J_1=0$,  $J_2\ne 0$, $J_9=0$ from (\ref{invT4}) then it is equaivalent to  $y''=\frac{y^{C+2}}{(C+1)(C+2)}+1$.

\item If $J_1=0$,  $J_2\ne 0$, $J_9\ne 0$, $K_1=0$ from (\ref{invT4}) then it is equaivalent to  $y''=\frac{y^{C+2}}{(C+1)(C+2)}+x$.

\item If $J_1=0$,  $J_2\ne 0$, $K_1\ne 0$ from (\ref{invT4}) then it is equaivalent to  $y''=\frac{y^{C+2}}{(C+1)(C+2)}+s(x),$ $s''(x)\ne 0.$

\noindent Two equations of Type IV.4 with the same parameter $C$ are equivalent if and only if after the transformation $\tilde x=J_2(x,y),\, \tilde y=J_9(x,y)$ their notations become identical.
 
\item If $J_1\ne 0$,  $J_2= 0$, $J_9\ne 0$, $K_1=0$ from (\ref{invT4}),  it is equaivalent to  $y''=\frac{y^{C+2}}{(C+1)(C+2)}+xy$.

\item If $J_1\ne 0$,  $J_2=0$,   $K=k=const\ne 0$ from (\ref{invT4}) then it is equaivalent to  $y''=\frac{y^{C+2}}{(C+1)(C+2)}+\frac{4y}{kx^2}$.

\noindent Two equations of Type IV.6 with the same parameter $C$ are equivalent if and only if invariant $K$ (\ref{invT4}) for both equations is the same and equal to the constant $k\ne 0$.

\item If $J_1\ne 0$,  $J_2\ne 0$,  $K=k=const\ne 0$ from (\ref{invT4}) then it is equaivalent to  $y''=\frac{y^{C+2}}{(C+1)(C+2)}+\frac{4y}{kx^2}+1$.

\noindent Two equations of Type IV.7 with the same parameter $C$ are equivalent if and only if invariant $K$ (\ref{invT4}) for both equations is the same and equal to the constant $k\ne 0$.

\item If $J_1\ne 0$,   $J_9= 0$, $K=0$, $J=a$ from (\ref{invT4}) then it is equaivalent to  

\noindent $y''=\frac{y^{C+2}}{(C+1)(C+2)}+y+a$.

\noindent Two equations of Type IV.8 with the same parameter $C$ are equivalent if and only if invariant $J$ (\ref{invT4}) for both equations is the same and equal to the constant $a$.

\item If $J_1\ne 0$,  $J_9\ne 0$,  $J=a=const$ from (\ref{invT4}) then it is equaivalent to  $y''=\frac{y^{C+2}}{(C+1)(C+2)}+t(x)y+at^{(C+2)/(C+1)}(x)$.

\noindent Two equations of Type IV.9 with the same parameter $C$ and invariant $J=a=const$ are equivalent if and only if after the transformation $\tilde x=J_1(x,y),\, \tilde y=J_9(x,y)$ their notations become identical.

\item If $J_1\ne 0$,  $J_2\ne 0$, $J_9\ne 0$, $K_1= 0$ from (\ref{invT4})  then it is equaivalent to  $y''=\frac{y^{C+2}}{(C+1)(C+2)}+xy+mx+n$.

\noindent Two equations of Type IV.10 with the equal parameter $C$ are equivalent if and only if in terms of canonical coordinates
their constants $m$ and $n$ are the same.
\item If $J_1\ne 0$,  $J_2\ne 0$, $J_9\ne 0$, $K_1\ne 0$, $K\ne const$, $J\ne const$ from (\ref{invT4}) then it is equaivalent to  $y''=\frac{y^{C+2}}{(C+1)(C+2)}+t(x)y+s(x)$.

\noindent Two equations of Type IV.11 with the same parameter $C$ are equivalent if and only if after the transformation $\tilde x=J(x,y),\, \tilde y=J_1(x,y)$ their notations become identical.

\end{enumerate}
In the cases IV.4, IV.9 and IV.11  $t(x)$ and $s(x)$ are definited up to
transformations (\ref{trTypeIV}).

\begin{table}[ht]
{Equations of Type IV.}
{\begin{tabular}{@{}cccccccl@{}} \hline
Type &  $J_1$ & $J_2$ & $ K$ & $J$ & $J_9$ & $K_1$ & Canonical form\\
 \hline
IV.1 &  $0$ &      $0$        & -                &     -       &  $0$       & $0$ &  $y''=\frac{y^{C+2}}{(C+1)(C+2)} $\\ 
IV.2 &  $0$ &      $\ne 0$    & -                &     -       &  $0$       & $0$ &  $y''=\frac{y^{C+2}}{(C+1)(C+2)}+1 $\\
IV.3 &  $0$ &      $\ne 0$    & -                &     -       &  $\ne 0$   & $0$ &  $y''=\frac{y^{C+2}}{(C+1)(C+2)}+x $\\
IV.4 &  $0$ &      $\ne 0$    & -                &     -       &  $\ne 0$   & $\ne 0$ &  $y''=\frac{y^{C+2}}{(C+1)(C+2)}+s(x) $\\
IV.5 &  $\ne 0$ &  $0$        & $\ne$const       &    0        &  $\ne 0$   & $0$ &  $y''=\frac{y^{C+2}}{(C+1)(C+2)} +xy$\\
IV.6 &  $\ne 0$ &  $0$        & $k=$const$\ne 0$ & 0           &  $\ne 0$   & $\ne 0$ &  $y''=\frac{y^{C+2}}{(C+1)(C+2)}+\frac{4y}{kx^2} $\\
IV.7 &  $\ne 0$ &  $\ne 0$    & $k=$const$\ne 0$ & $\ne$const &  $\ne 0$   & $\ne 0$ &  $y''=\frac{y^{C+2}}{(C+1)(C+2)}+\frac{4y}{kx^2}+1 $\\
IV.8 &  $\ne 0$ &  $\forall$  & 0                & $a=$const   &  $0$       & $0$ &  $y''=\frac{y^{C+2}}{(C+1)(C+2)} +y+a$\\
IV.9 &  $\ne 0$ &  $\forall$  & $\ne$const       & $a=$const   &  $\ne 0$   & $\forall$ &  $y''=\frac{y^{C+2}}{(C+1)(C+2)}+t(x)y+ $\\
& & & & & & & \phantom{....}$+at(x)^{\frac{C+2}{C+1}}$\\
IV.10 &  $\ne 0$ &  $\ne 0$   & $\ne$const       & $\ne$const  &  $\ne 0$   & $0$ &  $y''=\frac{y^{C+2}}{(C+1)(C+2)}+xy+$\\
& & & & & & & \phantom{....}$+mx+n $\\
IV.11 &  $\ne 0$ &  $\ne 0$   & $\ne$const       & $\ne$const  &  $\ne 0$   & $\ne 0$ &  $y''=\frac{y^{C+2}}{(C+1)(C+2)}+t(x)y+ $\\
& & & & & & & \phantom{....}$+s(x)$\\
\hline
\end{tabular}}
\end{table}
\end{theorem}

{\bf Proof.}  
Let us have the certain equation (\ref{eq}) of Type IV. Then in terms of canonical coordinates it has the form (\ref{typeIV}).

\begin{enumerate}
 \item If $J_1=0$ and $J_2=0$ then $t(x)\equiv 0$ and $s(x)\equiv 0$ and equation has the form 
$$y''=\frac{y^{C+2}}{(C+1)(C+2)}.$$

\item If $J_1=0$, $J_2\ne 0,$ $J_9=0$, then $t(x)\equiv  0$ and $s'(x)\equiv 0.$ So $s(x)=s=const\ne 0.$
Let's make the transformation (\ref{trTypeIV}). Then
$\tilde s = \alpha^{{2(C+2)}/{(C+5)}}\cdot s. $
Choosing the appropriate $\alpha$ we can make $\tilde s=1$,
therefore:
$$
\tilde y''=\frac{\tilde y^{C+2}}{(C+1)(C+2)}+1.
$$
\item If $J_1=0$, $J_2\ne 0,$ $J_9\ne 0$, $K_1=0$, then
$t(x)\equiv 0,$ $s''(x)\equiv 0$. So $s(x)=s_1x+s_2$, $s_1=const\ne 0,$ $s_2=const.$ 
Let's make the transformation (\ref{trTypeIV}), then
$$
\tilde s (\tilde x)= \alpha^{\frac{2(C+2)}{C+5}}\cdot (s_1(\alpha^{\frac{C+1}{C+5}} \tilde x+\beta)+s_2)=
\alpha^{\frac{3C+5}{C+5}}s_1\tilde x+\alpha^{\frac{2(C+2)}{C+5}}\cdot(s_1\beta+s_2).
$$
Choosing the parameters $\alpha$ and $\beta$ we can make $\tilde s (\tilde x)=\tilde x,$ then equation will be
$$
\tilde y''=\frac{\tilde y^{C+2}}{(C+1)(C+2)}+\tilde x.
$$

\item If $J_1=0$, $J_2\ne 0,$ $J_9\ne 0$ then $t(x)\equiv 0,$ $s''(x)\ne 0.$
Hence in terms of special coordinates:
$$
y''=\frac{y^{C+2}}{(C+1)(C+2)}+s(x).
$$

Let we have two equations of Type IV.4 with the same parameter $C$. 
The first way: we reduce both equations in the canonical forms
$$
y''=\frac{y^{C+2}}{(C+1)(C+2)}+s(x),\quad Y''=\frac{Y^{C+2}}{(C+1)(C+2)}+S(X).
$$
They equivalent if and only if exist the constants $\alpha$ and $\beta$ such that
$$
S(X)= \alpha^{\frac{2(C+2)}{C+5}}\cdot s(\alpha^{\frac{C+1}{C+5}} X+\beta).
$$

The second way:
note that for any equation (\ref{eq}) of the Type IV.4 invariants $J_2$ and $J_9$ are functionally independent. 
So,  we can  make the invariant point transformation
$$
\tilde x=J_2(x,y),\; \tilde y=J_9(x,y),\quad \tilde x=J_2(X,Y),\; \tilde y=J_9(X,Y)
$$
for both equations. Equations are equivalent if and only if in terms of new coordinates $(\tilde x,\tilde y)$ their
notations become identical.

\item  If $J_1\ne const $, $J_2=0,$ $J_9\ne const$, $K_1=0$, then $s(x)\equiv 0,$ $t''(x)\equiv 0,$ so $t(x)=t_1x+t_2,$ $t_1=const\ne 0$, $t_2=const.$ After the transformation (\ref{trTypeIV}):
$$
\tilde t(\tilde x) = \alpha^{\frac{2(C+1)}{C+5}}  \cdot (t_1(\alpha^{\frac{C+1}{C+5}} \tilde x+\beta)+t_2)=
\alpha^{\frac{3(C+1)}{C+5}}t_1 \tilde x+\alpha^{\frac{2(C+1)}{C+5}} \cdot(t_1\beta+t_2).
$$
Choosing parameters $\alpha$ and $\beta$ we can make $\tilde t (\tilde x)=\tilde x.$ Hence equation will be 
$$
\tilde y''=\frac{\tilde y^{C+2}}{(C+1)(C+2)}+\tilde x\tilde y.
$$

\item If $J_1\ne const $, $J_2=0,$ $J=0$, $K=k=const\ne 0$, hence
$s(x)\equiv 0$. In this case
$$
\frac{t^{\prime 2}(x)}{t^3(x)}=k,\quad t(x)=\frac{4}{(\sqrt k\cdot x+c_0)^2},\; c_0=const.
$$
After the transformation (\ref{trTypeIV}):
$$
\tilde t (\tilde x)=   \frac{4\alpha^{\frac{2(C+1)}{C+5}}}{(\sqrt k\cdot (\alpha^{\frac{C+1}{C+5}} \tilde x+\beta)+c_0)^2}  =
   \frac{4\alpha^{\frac{2(C+1)}{C+5}}}{(\alpha^{\frac{C+1}{C+5}}\sqrt k\cdot\tilde x+\beta\sqrt k+c_0)^2}.
$$
So  we can take the appropriate parameters $\alpha$ and $\beta$ so that $\tilde s=1$ and $\beta\sqrt k+c_0=0$:
$$
\tilde t(\tilde x)= \frac{4}{k\tilde x^2},\quad \tilde s=1,\quad 
\tilde y''=\frac{\tilde y^{C+2}}{(C+1)(C+2)}+\frac{4\tilde y}{k\tilde x^2}.
$$
Two equations of Type IV.6 with the same parameter $C$  are equivalent if and only if  invariants $K$ for both equations are the same and equal to constant $k\ne 0$.

\item If $J_1\ne const $, $J_2\ne const,$ $K=k=const\ne 0$, then
$s'(x)\equiv 0$ and $s(x)=s=const\ne 0$:
$$
\frac{t^{\prime 2}(x)}{t^3(x)}=k,\quad t(x)=\frac{4}{(\sqrt k\cdot x+c_0)^2},\; c_0=const.
$$
After the transformation (\ref{trTypeIV}): $\tilde s = \alpha^{\frac{2(C+2)}{C+5}}\cdot s,$
$$
\tilde t (\tilde x)=   \frac{4\alpha^{\frac{2(C+1)}{C+5}}}{(\sqrt k\cdot (\alpha^{\frac{C+1}{C+5}} \tilde x+\beta)+c_0)^2}  =
   \frac{4\alpha^{\frac{2(C+1)}{C+5}}}{(\alpha^{\frac{C+1}{C+5}}\sqrt k\cdot\tilde x+\beta\sqrt k+c_0)^2}.
$$
So  we can take the appropriate parameters $\alpha$ and $\beta$ so that $\tilde s=1$ and $\beta\sqrt k+c_0=0$: 
$$
\tilde t(\tilde x)= \frac{4}{k\tilde x^2},\quad \tilde s=1,\quad 
\tilde y''=\frac{\tilde y^{C+2}}{(C+1)(C+2)}+\frac{4\tilde y}{k\tilde x^2}+1.
$$
Two equations of Type IV.7 with the same parameter $C$  are equivalent if and only if  invariants $K$ for both equations are the same and equal to the constant $k\ne 0$.

\item If $J_1\ne const,$   $J_9=0$, then $t'(x)\equiv 0$ and $s'(x)\equiv 0$, so $t(x)=const\ne 0,$ $s(x)=const\ne 0.$
As $J=a=const$, then $s=at^{\frac{C+2}{C+1}}$.
Let us make the transformation (\ref{trTypeIV}), then
$$
\tilde t = \alpha^{\frac{2(C+1)}{C+5}}  \cdot t,\quad
\tilde s = \alpha^{\frac{2(C+2)}{C+5}}\cdot s.
$$
Choosing the appropriate $\alpha$ we can make $\tilde t=1$, then $\tilde s=a.$
In terms of new coordinates equation has the form
$$
\tilde y''=\frac{\tilde y^{C+2}}{(C+1)(C+2)}+\tilde y+a.
$$
Two equations of Type IV.8 with the same parameter $C$ are equivalent if and only if  invariants $J$ for both equations are the same and  equal to constant $a$.

\item  
If $J_1\ne 0,$   $J_9\ne 0$,  $J=a=const$, then $s(x)=at(x)^{\frac{C+2}{C+1}}$, $t(x)\ne const.$
Therefore in terms of special coordinates  equation has the form
$$
y''=\frac{y^{C+2}}{(C+1)(C+2)}+t(x)y+a\cdot t(x)^{\frac{C+2}{C+1}}.
$$
Let we have two equations of Type IV.9 with the same parameter $C$ and  invariant $J=a$. To solve the equivalence problem we can

1) reduce equations into the canonical forms
$$
y''=\frac{y^{C+2}}{(C+1)(C+2)}+t(x)y+a\cdot t(x)^{\frac{C+2}{C+1}},$$
$$
Y''=\frac{Y^{C+2}}{(C+1)(C+2)}+T(X)Y+a\cdot T(X)^{\frac{C+2}{C+1}}.
$$
They equivalent if and only if exist the constants $\alpha$ and $\beta$ such that
$$
T (X)= \alpha^{\frac{2(C+1)}{C+5}}  \cdot t(\alpha^{\frac{C+1}{C+5}} X+\beta).
$$

2) make the invariant point transformation. 
$$
\tilde x=J_1( x,\, y),\, \tilde y=J_9(x,\,  y),\quad 
\tilde x=J_1(X,\, Y),\, \tilde y=J_9(X,\, Y).
$$
Note that for any equation (\ref{eq}) of Type IV.9  invariants $J_1$ and  $J_9$ 
(\ref{invT4}) are functially independent. 
Equations are equivalent if and only if in terms of new coordinates $(\tilde x,\tilde y)$ their
notations become identical.

\item If $J_1\ne 0,$ $J_2\ne 0,$  $J_9\ne 0$,  $K_1=0$, then
$t''(x)\equiv 0$ and $s''(x)\equiv 0$. So $t(x)=t_1x+t_2$, $t_1=const\ne 0$, $t_2=const$; $s(x)=s_1x+s_2$, $s_1=const\ne 0$, $s_2=const$.
Let us make the transformation (\ref{trTypeIV}), then
$$
\tilde t(\tilde x) = \alpha^{\frac{2(C+1)}{C+5}}  \cdot (t_1(\alpha^{\frac{C+1}{C+5}} \tilde x+\beta)+t_2)=
\alpha^{\frac{3(C+1)}{C+5}}t_1 \tilde x+\alpha^{\frac{2(C+1)}{C+5}} \cdot(t_1\beta+t_2),
$$
$$
\tilde s (\tilde x)= \alpha^{\frac{2(C+2)}{C+5}}\cdot (s_1(\alpha^{\frac{C+1}{C+5}} \tilde x+\beta)+s_2)=
\alpha^{\frac{3C+5}{C+5}}s_1\tilde x+\alpha^{\frac{2(C+2)}{C+5}}\cdot(s_1\beta+s_2).
$$
Choosing the parameters $\alpha$ and $\beta$ we can make $\tilde t(\tilde x)=\tilde x$, but in this case
$\tilde s (\tilde x)=m\tilde x+n$, $m=const\ne 0$, $n=const.$ Thus  equation has the form
$$
\tilde y''=\frac{\tilde y^{C+2}}{(C+1)(C+2)}+\tilde x\tilde y+m\tilde x+n.
$$
Two equations of Type IV.9 with the equal parameter $C$ are equivalent if and only if in terms of canonical coordinates
their constants $m$ and $n$ are identical.

\item If $J_1\ne 0,$ $J_2\ne 0,$  $J_9\ne 0$, $K\ne const,$ $J\ne const$, $K_1\ne 0$, then 
in terms of special coordinates equation has the general form
$$
y''=\frac{y^{C+2}}{(C+1)(C+2)}+t(x)y+s(x).
$$
Let we have two equations of Type IV.11 with the same parameter $C$. To solve the equivalence problem we could

1) reduce both equations into the canonical forms
$$
y''=\frac{y^{C+2}}{(C+1)(C+2)}+t(x)y+s(x),$$
$$
Y''=\frac{Y^{C+2}}{(C+1)(C+2)}+T(X)Y+S(X).
$$
They equivalent if and only if exist the constants $\alpha$ and $\beta$ such that
$$
T (X)= \alpha^{\frac{2(C+1)}{C+5}}  \cdot t(\alpha^{\frac{C+1}{C+5}} X+\beta),\quad
S (X)= \alpha^{\frac{2(C+2)}{C+5}}\cdot s(\alpha^{\frac{C+1}{C+5}} X+\beta).
$$

2)  make the invariant point transformation. 
$$
\tilde x=J( x,\, y),\, \tilde y=J_1(x,\,  y),\quad 
\tilde x=J(X,\, Y),\, \tilde y=J_1(X,\, Y).
$$ Note that for any equation (\ref{eq}) of Type IV.11  invariants $J$ and  $J_1$ 
(\ref{invT4}) are functially independent. 
Equations are equivalent if and only if in terms of new coordinates $(\tilde x,\tilde y)$ their
notations become identical.
\end{enumerate}

\subsection{Examples}

\subsubsection{Equation Painleve III with 3 zero parameters}

The  Painleve III equation depends on 4 parameters $(a,b,c,d)$
$$
PIII(a,b,c,d):\qquad\qquad y''=\frac{y^{\prime 2}}{y}-\frac{y'}{x}+\frac{(ay^2+b)}{x}+cy^3+\frac{d}{y}.
$$
By the paper \cite{Hietarinta} all equations Painleve III with 3 zero parameters
are equivalent
$$
PIII(0,b,0,0)\stackrel{1)}{\rightarrow}PIII(-b,0,0,0)\stackrel{2)}{\rightarrow}PIII(0,0,-b,0)\stackrel{3)}{\rightarrow}PIII(0,0,0,b),
$$
where the point transformations 1) and 3) are: $x=\tilde x,\,y=1/{\tilde y}$ and 2) is: $x=\tilde x^2/2,\,y=\tilde y^2.$

 In this way suppose that  $a\ne 0$, $b=c=d=0.$  For the equation $PIII(a,0,0,0)$
 conditions (\ref{cond}) are hold and  invariants (\ref{inv1}) are equal to:
$$
I_1=\frac{3}{5},\quad I_2=0,\quad I_3=\frac{1}{15}.
$$
As $J_3=0$, then according to Theorem \ref{the2}, equation $PIII(a,0,0,0)$ has Type I.1.
Let us find the corresponding change of variables.
The first point transformation (see \cite{Babich}) takes equation $PIII(a,0,0,0)$ into the 
following form:
$x=e^t,$ $ y=e^z, $ $ z''=ae^{t+z}.$

The second transformation reduces it in the canonical form: 
$
t={\tilde x}/{\sqrt a},$ $ z=\tilde y-{\tilde x}/{\sqrt a},$ $
\tilde y''=e^{\tilde y}.$

\subsubsection{Equation from the handbook Kamke No. 6.172}
Equation from the handbook Kamke \cite{Kamke} No. 172
$$
y''=\frac{y^{\prime 2}}{y}-\frac{ay'}{x}-by^2,\quad b\ne 0
$$
has  Type I and  invariants (\ref{inv1}), (\ref{inv2}):
$$
I_1=\frac{3}{5},\; I_2=0,\; I_3=\frac{1}{15}+\frac{2a}{15(a-1)x^2 e^y},\;
 I_6=\frac{2a}{5(a-1)x^2e^y},\;
I_9=\frac{16a^2}{1875(a-1)^2x^6e^{3y}}.
$$
Let us calculate the additional invariants (\ref{invT1}):
$$
J_3=J_6=-\frac{2a(a-1)}{ybx^2},\quad  K=\frac{2(a-1)}{a}.
$$
According to Theorem \ref{the2} this equation has Type I.4. with $k=2(a-1)/a$ if $a\ne 0$ and $a\ne 1$. 
In the special cases $a=0$ or $a=1$ equation has Type I.1.

Let us find corresponding change of variables.
The first transformation (as $b\ne 0$):
$$
x=\frac{t}{\sqrt{-b}},\quad y=z,\quad 
\tilde z''=\frac{ z^{\prime 2}}{z}-\frac{a z'}{t}+ z^2.
$$
Then, as $a\ne 1$,  the second transformation
$$
t=(1-a)^{\frac{1}{1-a}} X^{\frac{1}{1-a}},\quad z=e^{Y},\quad 
Y''=(1-a)^{\frac{2a}{1-a}}{X}^{\frac{2a}{1-a}}e^{Y}.
$$
In the end, as $a\ne 0$,  the third transformation:
$$
X=\tilde x,\quad Y=\tilde y+\frac{2a}{a-1}\ln((1-a)\tilde x),\quad
\tilde y''=\frac{2a}{(a-1)\tilde x^2}+e^{\tilde y}.
$$
If $a=1$, then the second and the third transformations:
$$
t =e^{X},\quad z=e^{Y},\quad Y''=be^{Y+2X};\qquad
X=\tilde x,\quad Y=\tilde y-2\tilde x,\quad \tilde y''=e^{\tilde y}.
$$

\subsubsection{Equation Painleve II}
The  Painleve II equation depends on  one parameter $a$ 
$$
PII(a):\qquad\qquad y''=2y^3+xy+a.
$$
It  also has form (\ref{model}) and
 satisfies to the conditions (\ref{cond}) with invariant $I_1={18}/{5}$. According to Theorem \ref{the5} 
it has Type IV, case $C=1.$
Let us calculate the additional invariants
$$
J_1=\frac{x}{12y^2},\quad J_2=\frac{a}{12y^3},\quad J=\frac{2a\sqrt 3}{x\sqrt x},\quad
K=\frac{1}{x^3},\quad J_9=\frac{1}{1728y^6},\quad K_1=0.
$$
We see that the equation PII(a) has Type IV.10  if $a\ne 0$ and Type IV.5 if $a=0$.

Let us calculate  $x,$ $y$ and $a$ via invariants:
$x= 1/{\sqrt[3]{K}}$, $y={1}/{(2\sqrt 3\sqrt[6]{J_9})}$, $a={J_2}/{2\sqrt 3\sqrt{J_9}}.$ 

\begin{theorem}\label{the6} Equation (\ref{eq}) of Type IV is equivalent to Painleve II equation with parameter $\pm a $  if and only if
\begin{equation}\label{p2a}
C=1,\quad \frac{J_2}{2\sqrt 3\sqrt{J_9}}=a=const,\quad K_1=0.
\end{equation}
The explicit point transformation is 
$\tilde x= 1/{\sqrt[3]{K(x, y)}},$  $\tilde y={1}/{(2\sqrt 3\sqrt[6]{J_9(x,y)})}.$
\end{theorem}

{\bf Proof.}
At first we write the PII(a) equation in the canonical form:
$$
x=X,\quad y=\frac{Y}{2\sqrt 3}\quad Y''=\frac{Y^3}{6}+XY+2\sqrt 3 a.
$$

Let $a=0$, then  equation PII(0) is already of Type IV.5. 

Let $a\ne 0$.  If for the certain equation of Type IV.10  conditions (\ref{p2a}) are hold, then in terms of
canonical coordinates 
$$
\frac{s(x)y^{\frac{C+1}{2}}}{t'(x)y+s'(x)}=2\sqrt 3 a.
$$ 
It is true if and only if $C=1$ and $s'(x)\equiv 0$. So $s(x)=s=const\ne 0.$  Then $t(x)={sx}/{(2\sqrt 3a)}+t,$ $t=const$, therefore $K_1=0.$

Let us make the point transformation (\ref{trTypeIV}):
$$
\tilde t (\tilde x)= \alpha^{\frac{2}{3}} \left( \frac {s(\alpha^{\frac{1}{3}} \tilde x+\beta)}{2\sqrt 3a}+t \right),\quad
\tilde s = \alpha \cdot s.
$$
If we take $\alpha=2\sqrt 3a/s$, $\beta=-ts^{\frac{2}{3}}(2\sqrt 3 a)^{\frac{1}{3}}$, then
$\tilde t(\tilde x)=\tilde x,$ $ \tilde s=2\sqrt 3 a $
and we get the Painleve II equation that is written in terms of the canonical coordinates $(\tilde x,\, \tilde y)$.

\subsubsection{Equation from the handbook Kamke No. 6.24}

Equation from the handbook Kamke \cite{Kamke} No. 6.24
$$
y''=-3ay'+2y^3-2a^2y
$$
has Type IV and  invariants $I_1={18}/{5}$, $I_3={1}/{15}$, $J_1=0,$ $J_2=0.$
According to Theorem   \ref{the5}  this equation has Type IV.1 with $C=1$. Let us make the 
transformations:
$$
x=-\frac{\ln(-3a X)}{3a},\quad y=Y,\quad 
Y''=-\frac{2Y(a^2-Y^2)}{9a^2X^2};$$
$$
X=\frac{a^2\tilde x^3}{36},\quad Y=\frac{a\tilde x\tilde y}{2\sqrt{3}},\quad \tilde y''=\frac{\tilde y^3}{6}.
$$

\subsubsection{Equation from the handbook Kamke  No. 6.140}

Equation from the handbook Kamke \cite{Kamke}  No. 6.140
$$
y''=\frac{y'^2}{2y}+4y^2
$$
has Type IV and invariants $I_1={18}/{5}$, $I_3={1}/{15},$ $J_1=0,$ $J_2=0.$
Thus it has Type IV.1 with $C=1$. By the  transformations $x={\tilde x}/{\sqrt{3}},$ $y={\tilde y^2}/{4}$ it is reduced into the canonical form:
 $ \tilde y''={\tilde y^3}/{6}.$

So equations from the handbook Kamke No. 6.24 and  No. 6.140 are equivalent.

\subsubsection{Equation from the handbook Kamke  No. 6.141}

Equation from the handbook Kamke \cite{Kamke}  No. 6.141
$$
y''=\frac{y'^2}{2y}+4y^2+2y
$$
has Type IV and invariants $I_1={18}/{5}$, $J_1={1}/{(12y)}$, $J_2=J=K=J_9=0.$
Hence it has Type IV.8  with $a=0$.
Thus it can be reduced into the canonical form: 
$$ x=\tilde x,\quad y=\frac{\tilde y^2}{12},\qquad \tilde y''=\frac{\tilde y^3}{6}+\tilde y.$$

\subsubsection{Emden equation}
The Emden equation in the form described in paper \cite{BordagBandle}:
\begin{equation}\label{emden}
y''=\lambda(x) y^n,\quad n\ne -3,\,0,\, 1,\, 2,\quad \lambda(x)\ne 0
\end{equation}
has $I_1={3(n+3)}/{(5(n-2))},$ $I_2=0,$ $ J_2=0$ hence 
this equation has Type IV with parameter $C=n-2$. Possible cases IV.1, IV.5, IV.6, IV.8 and IV.9.

Invariants (\ref{invT4}) are the follows:
$$
\aligned
J_1=&\frac{\lambda\lambda ''(n+3)-\lambda^{\prime 2}(n+4)}{(n+3)^2(n-1)n\lambda^3y^{n-1}},\\
J_9=&\frac{((n+3)^2\lambda '''\lambda^2-3(n+3)(n+5)\lambda '\lambda ''+ 2(n+4)(n+5)\lambda^{\prime 3})^2}{(n+3)^6(n-1)^3n^3\lambda(x)^9y^{3(n-1)}},\\
K=&\frac{J_9}{J_1^3}=\frac{((n+3)^2\lambda '''\lambda^2-3(n+3)(n+5)\lambda '\lambda ''+ 2(n+4)(n+5)\lambda^{\prime 3})^2}{(\lambda(x)\lambda ''(x)(n+3)-\lambda (x)^{\prime 2}(n+4))^3},\\
K_1=&\frac{{K_1^*}}{(n+3)^4(n-1)^2n^2\lambda(x)^{6}y^{2(n-1)}},\\
&K_1^*=
(n+3)^3\lambda^3\lambda ''''-4(n+3)^2(n+6)\lambda^2\lambda '\lambda '''-3(n+3)^2(n+5)\lambda^2\lambda^{\prime\prime 2}+\\
&+12(n+3)(n+5)^2\lambda\lambda^{\prime 2}\lambda ''-6(n+4)(n+5)^2\lambda^{\prime 4}=0.
\endaligned
$$

\noindent {\bf Type IV.1.} The Emden equation (\ref{emden}) has Type IV.1 if and only if invariant $J_1=0$. 

In this case  function $\lambda (x)$ is the solution of following differential equation:
\begin{equation}\label{IVJ1}
 \lambda\lambda ''(n+3)=\lambda^{\prime 2}(n+4).
\end{equation}
Therefore $\lambda(x)={c_2}/{(c_1x+1)^{n+3}},$ $c_1,\, c_2=const,$ $c_2\ne 0.$
In this case the Emden equation could be transformed into the canonical form 
$$
\aligned
\tilde y''=\frac{\tilde y^n}{(n-1)n},\qquad
&x=-\frac{\sqrt[n+1]{n(n-1)c_2}}{\tilde x c_1\sqrt[n+1]{c_1^2}}-\frac{1}{c_1},\quad y=-\frac{\tilde y}{\tilde x},\quad c_1\ne 0,\\
&x=\frac{\tilde x}{\sqrt{c_2n(n-1)}},\quad\qquad\qquad y=\tilde y,\qquad\; c_1=0.
\endaligned
$$

\noindent {\bf Type IV.8.} The Emden equation (\ref{emden}) has Type IV.8   if and only if  invariants $J_1\ne 0$, $J_9=0$. Then $\lambda (x)$ satisfies to the following differential equation
\begin{equation}\label{IVJ9}
(n+3)^2\lambda '''\lambda^2-3(n+3)(n+5)\lambda '\lambda ''+ 2(n+4)(n+5)\lambda^{\prime 3}=0
\end{equation}
and doesn't satisfies to  (\ref{IVJ1}). Then  equation (\ref{emden}) 
could be transformed into 
$$
\tilde y''=\frac{\tilde y^n}{(n-1)n}+\tilde y.
$$

\noindent {\bf Type IV.5.} The Emden equation (\ref{emden}) has Type IV.5  if and only if  invariants $J_1\ne 0$, $J_9\ne 0$, $K\ne const$, $K_1=0$. Then $\lambda (x)$  satisfies to the following  differential equation
\begin{equation}\label{IVK1}
\aligned
&(n+3)^3\lambda^3\lambda ''''-4(n+3)^2(n+6)\lambda^2\lambda '\lambda '''-3(n+3)^2(n+5)\lambda^2\lambda^{\prime\prime 2}+\\
&+12(n+3)(n+5)^2\lambda\lambda^{\prime 2}\lambda ''-6(n+4)(n+5)^2\lambda^{\prime 4}=0
\endaligned
\end{equation}
and doesn't satisfies to  (\ref{IVJ1}), (\ref{IVJ9}), (\ref{IVK}). Equation (\ref{emden}) could be transformed into 
$$
\tilde y''=\frac{\tilde y^n}{(n-1)n}+\tilde x\tilde y.
$$

In particular, this equations is equivalent to the Painleve II with the parameter $a=0$
if and only if  $n=3$ and the function $\lambda(x)$  satisfies to  the following equation:
$$
9\lambda^3\lambda ''''-54\lambda^2\lambda '\lambda '''-36\lambda^2\lambda^{\prime\prime 2}+192\lambda\lambda^{\prime 2}\lambda ''
-112\lambda^{\prime 4}=0.
$$

\noindent {\bf Type IV.6.} The Emden equation (\ref{emden}) has Type IV.6  if and only if  invariants $J_1\ne 0$, $J_9\ne 0$, $K=k=const\ne 0$, $K_1\ne 0$. Then $\lambda (x)$  satisfies to the following  equation
\begin{equation}\label{IVK}
\aligned
&\lambda^{\prime\prime\prime 2}\lambda^4(n+3)^4-
2\lambda^{\prime\prime\prime }\lambda '\lambda^2(n+5)(n+3)^2\left( 3\lambda ''\lambda (n+3)-2\lambda^{\prime 2}(n+4) \right)-\\
&-\lambda^{\prime\prime 3}\lambda^3 k(n+3)^3+
3\lambda^{\prime\prime 2}\lambda^{\prime 2}\lambda^2(n+3)^2(3n^2+30n+kn+75+4k)-\\
&-3\lambda^{\prime\prime }\lambda^{\prime 4}\lambda(n+4)(n+3)(4n^2+40n+kn+100+4k)+\\
&+\lambda^{\prime 6}(n+4)^2(4n^2+40n+kn+100+4k)=0
\endaligned
\end{equation}
and doesn't satisfies to  (\ref{IVJ1}), (\ref{IVJ9}), (\ref{IVK1}).
Equation (\ref{emden}) could be transformed into 
$$
\tilde y''=\frac{\tilde y^n}{(n-1)n}+\frac{4\tilde y}{k\tilde x^2}.
$$

\noindent {\bf Type IV.9.}
At the other cases the Emden equation (\ref{emden}) has Type IV.9. Then exists the special coordinate system, so that in terms of these coordinates equation has form
$$
\tilde y''=\frac{\tilde y^n}{(n-1)n}+\tilde t(\tilde x)\tilde y.
$$

\section*{Acknowledgments}

I am grateful to Prof.\ Dr.\ Rainer Picard and the Institute of Analysis of the Technische Universit\"at Dresden where this work was done.
Also many thanks to my colleagues from Ufa, Russia: to Prof.\ B.I.Suleimanov for his useful remarks and to Dr.\ R.N.Garifullin for his help in implementation of the Buchberger algorithm.

The work supported by the German Academic Exchange Service (DAAD), programme ``Mikhail Lomonosov II'' 2010, coordinator Prof.\ Rainer Picard,  Institute for Analysis, Technische Universit\"at Dresden. 

The work partially supported by the {R}ussian 
{F}ond of {B}asic {R}esearch (RFBR), grant  08-01-97020-p\_povoljie\_a, coordinator Prof. Fazullin Z.Yu., Math. Department, Bashkir State University, Ufa, Russia.



\appendix
\section{Explicit formulas for the components of the pseudovectorial fields, pseudoinvariants and invariants}
Here and everywhere the notation $K_{i.j}$ denotes partial differentiation: $K_{i.j}={\partial ^{i+j}K}/{\partial x^i\partial y^j}.$ 

The coordinates of the pseudovectorial field ${ \alpha}$ are
$\alpha^1=B$, $\alpha^2=-A$,
where
\begin{equation}\label{alpha}
\aligned A=P_{ 0.2}&-2Q_{ 1.1}+R_{ 2.0}+ 2PS_{ 1.0}+SP_{
1.0}-3PR_{ 0.1}-3RP_{ 0.1} -3QR_{ 1.0} +6QQ_{ 0.1}, \\
 B=S_{ 2.0}&-2R_{ 1.1}+Q_{ 0.2}- 2SP_{0.1}-
PS_{ 0.1}+3SQ_{ 1.0}+3QS_{ 1.0}+ 3RQ_{ 0.1}-6RR_{ 1.0}.
\endaligned
\end{equation}

The pseudoinvariant $F$ of  weight 5 is:
\begin{equation}\label{F}
3F^5=AG+BH,\qquad\text{where}
\end{equation}
$$ \aligned G&=-BB_{ 1.0}-3AB_{ 0.1}+4BA_{ 0.1}+
3SA^2-6RBA+3QB^2,\\
H&=-AA_{ 0.1}-3BA_{ 1.0}+4AB_{ 1.0}-
3PB^2+6QAB-3RA^2.
\endaligned
$$

The pseudoinvariant $N$ in the cases $A\ne 0$ and $B\ne 0$, respectively, is: 
\begin{equation}\label{N}
 N =-\frac H{3A}, \qquad\qquad N =\frac G{3B}.
\end{equation}

The pseudoinvariant $M$  in the case $A\ne 0$:
\begin{equation}\label{M1}
M=-\frac {12BN(BP+A_{ 1.0})}{5A}+BN_{ 1.0}+\frac {24}5BNQ+\frac
65NB_{ 1.0}+\frac 65NA_{ 0.1}-AN_{ 0.1}- \frac {12}5ANR.
\end{equation}

And in the case $B\ne 0$ is:
\begin{equation}\label{M2}
M=-\frac {12AN(AS-B_{ 0.1})}{5B}-AN_{ 0.1}+\frac {24}5ANR-
 \frac
65NA_{ 0.1}-\frac 65NB_{ 1.0}+BN_{ 1.0}- \frac {12}5 BNQ.
\end{equation}

The pseudoinvariant  $\Omega$ in the case $A\ne 0$:
\begin{equation}\label{Omega1}
\aligned \Omega &=\frac {2BA_{ 1.0}(BP+ A_{ 1.0})}{A^3}- \frac
{(2B_{ 1.0}+3BQ)A_{ 1.0}}{A^2}+\frac {(A_{ 0.1}-2B_{1.0})BP}{A^2}-\\
&- \frac {BA_{ 2.0}+B^2 P_{ 1.0}}{A^2}+ \frac
{B_{ 2.0}}A+\frac {3B_{ 1.0}Q+3BQ_{ 1.0}- B_{ 0.1}P-BP_{
0.1}}{A}+Q_{ 0.1}- 2R_{ 1.0}.
\endaligned
\end{equation}

The pseudoinvariant  $\Omega$ in the case  $B\ne 0$: 
\begin{equation}\label{Omega2}
\aligned \Omega &=\frac {2AB_{ 0.1}(AS- B_{ 0.1})}{B^3}- \frac
{(2A_{ 0.1}-3AR)B_{ 0.1}}{B^2}+\frac {(B_{ 1.0}-2A_{
0.1})AS}{B^2}+ \\
&+\frac {AB_{ 0.2}-A^2 S_{ 0.1}}{B^2}- \frac
{A_{ 0.2}}B+\frac {3A_{ 0.1}R+3AR_{ 0.1}- A_{ 1.0}S-AS_{
1.0}}{B}+R_{ 1.0}- 2Q_{ 0.1}.
\endaligned
\end{equation}

In the case $A\ne 0$ the field ${ \gamma}$ is:
\begin{equation}\label{gamma1}
\aligned
 \gamma^1=&-\frac {6BN(BP+A_{ 1.0})}{5A^2}+
\frac {18NBQ}{5A}+
\frac {6N(B_{ 1.0}+A_{ 0.1})}{5A} -N_{ 0.1}-\frac
{12}5NR-2\Omega B,\\
\gamma^2=&-\frac {6N(BP+A_{ 1.0})}{5A}+N_{ 1.0}+\frac 65NQ+ 2\Omega
A.
\endaligned
\end{equation}

In the case $B\ne 0$ the field ${ \gamma}$ is:
\begin{equation}\label{gamma2}
\aligned
\gamma^1=&-\frac {6N(AN-B_{ 0.1})}{5B}-N_{ 0.1} +\frac 65NR-2\Omega
B,\\
 \gamma^2=&-\frac {6AN(AS-B_{ 0.1})}{5B^2}+
\frac {18NAR}{5B}-
\frac {6N(A_{ 0.1}+B_{ 1.0})}{5B} +N_{ 1.0}-\frac
{12}5NQ+2\Omega A.
\endaligned
\end{equation}

The pseudoinvariant  $\Gamma$ is:
\begin{equation}\label{Gamma}
\aligned
\Gamma=&\frac {\gamma^1\gamma^2(\gamma^1_{
1.0}- \gamma^2_{ 0.1})}{M}+ \frac {(\gamma^2)^2\gamma^1_{ 0.1}-
(\gamma^1)^2\gamma^2_{ 1.0}}M+\\
&+\frac
{P(\gamma^1)^3+3Q(\gamma^1)^2\gamma^2+3R\gamma^1(\gamma^2)^2+
S(\gamma^2)^3}M.
\endaligned
\end{equation}

Explicit formulas for the invariants $I_6$, $I_9$, $I_{21}$:
\begin{equation}\label{inv6921}
I_6=\frac{(I_3)'_xB-(I_3)'_yA}{N},\quad I_9=\frac{\left((I_3)'_x\gamma^1+(I_3)'_y\gamma^2\right)^2}{N^3},\quad
I_{21}=\frac{\left((I_9)'_x\gamma^1+(I_9)'_y\gamma^2\right)^2}{N^3}.
\end{equation}

\end{document}